\documentclass[aps,pra,twocolumn,amsmath,amssymb,showpacs,superscriptaddress,notitlepage,longbibliography]{revtex4-1}
\usepackage[colorlinks=true,linkcolor=blue,anchorcolor=red,citecolor=blue, urlcolor=blue]{hyperref}
\usepackage{bm}
\usepackage{graphicx}
\usepackage{color}

\begin{document}
\title{Performance of the T-matrix based master equation for Coulomb drag in double quantum dots}

\author{Wan-Xiu He}
\affiliation{School of Physical Science and Technology $\&$ Key Laboratory for Magnetism and Magnetic Materials of the Ministry of Education, Lanzhou University, Lanzhou 730000, China}

\author{Zhan Cao}
\email{caozhan@baqis.ac.cn}
\affiliation{Beijing Academy of Quantum Information Sciences, Beijing 100193, China}

\author{Gao-Yang Li}
\affiliation{School of Physical Science and Technology $\&$ Key Laboratory for Magnetism and Magnetic Materials of the Ministry of Education, Lanzhou University, Lanzhou 730000, China}

\author{Lin Li}
\affiliation{College of Physics and Electronic Engineering, and Center for Computational Sciences, Sichuan Normal University, Chengdu, 610068, China}

\author{Hai-Feng L\"{u}}
\affiliation{School of Physics, University of Electronic Science and Technology of China, Chengdu 610054, China}

\author{ZhenHua Li}
\affiliation{School of Physical Science and Technology $\&$ Key Laboratory for Magnetism and Magnetic Materials of the Ministry of Education, Lanzhou University, Lanzhou 730000, China}

\author{Hong-Gang Luo}
\email{luohg@lzu.edu.cn}
\affiliation{School of Physical Science and Technology $\&$ Key Laboratory for Magnetism and Magnetic Materials of the Ministry of Education, Lanzhou University, Lanzhou 730000, China}
\affiliation{Beijing Computational Science Research Center, Beijing 100193, China}

\begin{abstract}
Recently, novel Coulomb drag mechanisms in capacitively coupled double quantum dots were uncovered by the T-matrix based master equation (TME). The TME is so far the primary approach to studying Coulomb drag in the weak-coupling regime; however, its accuracy and reliability remain unexplored. Here, we evaluate the performance of the TME for Coulomb drag via a comparison with numerically exact results obtained by the hierarchical equation-of-motion approach. We find that the TME can capture qualitative current evolutions versus dot levels, temperature, and effective coupling strengths, but only partially succeeds at the quantitative level. Specifically, the TME gives highly inaccurate drag currents when large charge fluctuations on dots exist and the fourth-order tunneling processes make a leading-order contribution. This failure of the TME is attributed to the combined effect of the unique drag mechanisms and its overlook of the fourth-order single-electron tunnelings. We identify the reliable regions to facilitate further quantitative studies on Coulomb drag by the TME.
\end{abstract}

\maketitle

\section{Introduction}\label{int}
The phenomenon that a current in one driven conductor induces a current or voltage in a nearby undriven conductor via Coulomb interactions is dubbed Coulomb drag \cite{RevModPhys.88.025003}. Since the early 1990s, Coulomb drag has been widely observed in low-dimensional semiconductor structures such as parallel electron-gas layers \cite{PhysRevLett.63.2508,PhysRevLett.66.1216,PhysRevB.47.12957,gramila1994measuring,PhysRevLett.68.1196}, quantum wires \cite{debray2000experimental,debray2001experimental,yamamoto2006negative}, quantum point contacts \cite{PhysRevLett.99.096803}, and quantum dots \cite{shinkai2009bidirectional,PhysRevLett.84.1986,PhysRevLett.96.176601}. Recent progress in material fabrication has fueled the interest in Coulomb drag in diverse graphene-based devices \cite{PhysRevB.83.161401,gorbachev2012strong,kim2012coulomb,chen2013coulomb,gamucci2014anomalous,PhysRevLett.117.046803,PhysRevLett.122.186602} and topological materials \cite{carrega2012theory,PhysRevB.88.235420,PhysRevLett.115.186404,liu2016coulomb,PhysRevB.95.075141,PhysRevB.95.205435}. Generally, a drag current is induced by Coulomb-mediated momentum and/or energy transfer between the carriers in the drive and drag systems. Nonetheless, the investigation of the fundamental details influencing the drag current is still in progress \cite{PhysRevB.96.075305,PhysRevB.99.035423,PhysRevB.99.165404}.

Recently, the Coulomb drag in a capacitively coupled double quantum dot comprising two stacked graphene nanoribbons was experimentally studied \cite{bischoff2015measurement}. The drag current was measurable in the forbidden region predicted by sequential-tunneling-only drag \cite{PhysRevLett.104.076801}, which triggered a renewed effort to explore the drag mechanism underneath. By exploiting the T-matrix based master equation (TME) approach \cite{bruus2004many} involving the fourth-order tunneling processes, it was established that cotunneling-assisted and cotunneling-only drag mechanisms are crucial for understanding the drag behavior \cite{PhysRevLett.116.196801,PhysRevLett.117.066602,lim2018engineering}. Cotunneling processes were fully respected in subsequent studies on thermoelectrical Coulomb drag \cite{thierschmann2016thermoelectrics,PhysRevB.96.115414,PhysRevB.96.115415,PhysRevB.98.035415}. These advances were achieved in the weak-coupling regime where the TME was believed to be reliable. Due to the high efficiency and convenience, the TME has become the primary approach to studying Coulomb drag in quantum dot systems, but without an explicit check of its accuracy and reliability. Considerable quantitative inconsistency between the output power of a Coulomb drag based thermal engine obtained by the TME and the noncrossing approximation was implied recently \cite{PhysRevB.96.115414}. Moreover, the prior application of the TME to the single-impurity Anderson model with a magnetic field demonstrated that the TME basically works well only in the deep Coulomb blockade regime \cite{PhysRevB.82.235307}. The reason is that the TME overlooks some fourth-order tunneling processes that are negligible inside the deep Coulomb blockade regime but relevant in other regimes \cite{PhysRevB.82.235307,PhysRevB.82.045316}. Given the unique Coulomb drag mechanisms and the deficiency of the TME, a detailed examination of whether and to what extent the TME can provide a quantitatively reliable drag current in quantum dot systems is required.

In this work, we evaluate the performance of the TME for Coulomb drag in capacitively coupled double quantum dots in the weak-coupling regime. To this end, we compare the results obtained by the TME and the numerically exact hierarchical equation of motion (HEOM) \cite{jin2008exact,zheng2009numerical,PhysRevLett.109.266403}, the latter is applicable to transport through quantum dots with Coulomb interactions. We find that the TME can capture qualitative current evolutions versus dot levels, temperature, and effective coupling strengths. However, at the quantitative level, the TME has quite different performances in the regions dominated by different drag mechanisms. The TME succeeds in obtaining a quantitatively satisfying drag current in most cases. However, it generally fails in a wide region dominated by cotunneling-assisted drag and in a narrow region dominated by sequential-tunneling-only drag. In both situations, large charge fluctuations on dots exist, and the fourth-order tunneling processes make a leading-order contribution to the drag current. Such a failure of the TME is attributed to the fact that the fourth-order single-electron tunnelings (SETs) possessing an intermediate charge fluctuation on dots are completely overlooked. This deficiency of the TME is not serious for conventional transport directly driven by a voltage or temperature bias, as the fourth-order SETs are either suppressed or make a next-to-leading order correction to the current. By contrast, due to the unique Coulomb drag mechanisms, the fourth-order SETs may make a leading-order contribution to the drag current comparable to the other fourth-order tunnelings captured by the TME, resulting in a highly inaccurate drag current.

The remainder of this paper is organized as follows. We present the model Hamiltonian and necessary details of the TME and HEOM approaches in Sec.~\ref{ModelandApp}. Numerical results and discussion are presented in Sec.~\ref{results}. Finally, we give a summary and outlook in Sec.~\ref{Summary}.

\begin{figure}[t!]
\centering
\includegraphics[width=\columnwidth]{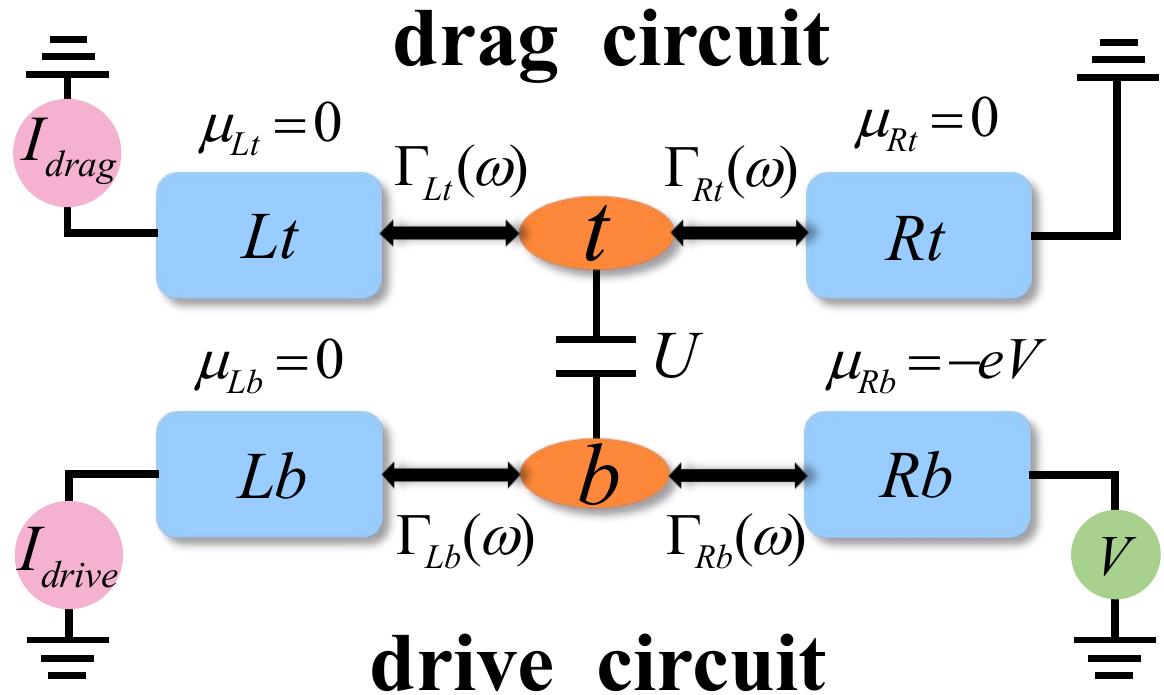}
\caption{Schematic of a capacitively coupled double quantum dot with an interdot Coulomb interaction $U$. The top ($t$) and bottom ($b$) dots are tunnel coupled to their nearest left ($L$) and right ($R$) leads with energy-dependent hybridizations $\Gamma(\omega)$. A bias voltage $V$ is applied to lead $Rb$ while the other leads are grounded. A current driven by the bias voltage in the bottom circuit would drag a directional current in the top circuit due to the interplay of $U$ and $\Gamma(\omega)$, which is dubbed Coulomb drag. }\label{Fig:Model}
\end{figure}

\section{Model and approaches}\label{ModelandApp}
\subsection{Model}\label{model}
The capacitively coupled double quantum dots in experiments \cite{bischoff2015measurement,PhysRevLett.117.066602} are depicted in Fig.~\ref{Fig:Model} and can be modeled by $H=H_{\textrm{DQD}}+\sum_{\alpha m} H_{\alpha m}+H_{T}$,
with \cite{PhysRevLett.104.076801,PhysRevLett.116.196801,PhysRevLett.117.066602}
\begin{eqnarray}
H_{DQD}&=&\sum_{m=t,b}\varepsilon_{m}d^\dag_{m}d_{m}+Ud^\dag_td_t d^\dag_b d_b,  \\
H_{\alpha m}&=&\sum_{k}(\varepsilon_{k\alpha m}-\mu_{\alpha m})c^\dag_{k\alpha m}c_{k\alpha m},  \\
H_{T}&=&\sum_{k,\alpha,m}(t_{k\alpha m}c^\dag_{k\alpha m}d_m+\textrm{H.c.}),\label{HT}
\end{eqnarray}
where $d^\dag_m$ ($c^\dag_{k\alpha m})$ creates an electron with energy $\varepsilon_m$ ($\varepsilon_{k\alpha m})$ on the dot (lead) in the top ($m=t$) or bottom ($m=b$) circuit. $H_{DQD}$ describes a double quantum dot with an interdot Coulomb interaction $U=e^2/2C$ arising from the effective capacitance $C$ between the dots. $H_{\alpha m}$ with $\alpha m=\{Lt, Lb, Rt, Rb\}$ models the noninteracting metallic lead with the chemical potential $\mu_{\alpha m}$. As indicated in Fig.~\ref{Fig:Model}, we treat the top (bottom) circuit as the drag (drive) circuit by setting $\mu_{Lt}=\mu_{Rt}=\mu_{Lb}=0$ and $\mu_{Rb}\ne 0$. $H_T$ describes tunnel couplings between each dot and its nearest left ($\alpha=L$) and right ($\alpha=R$) leads with $t_{k\alpha m}$ being the tunneling matrix element. The hybridization between dot $m$ and lead $\alpha m$ is defined as $\Gamma_{\alpha m}(\omega)\equiv 2\pi\sum_k\vert t_{k\alpha m}\vert^2 \delta(\omega-\varepsilon_{k\alpha m})$, which depends on both the single-particle dispersion of the lead electron and the property of the tunnel barrier.

As previous studies \cite{PhysRevLett.104.076801,PhysRevLett.116.196801,PhysRevLett.117.066602} pointed out, to engineer a drag current, apart from a Coulomb $U$, energy-dependent hybridizations satisfying $\Gamma_{Lt}(\omega)\ne k\Gamma_{Rt}(\omega)$, with $k$ being a constant, are essential in the drag circuit. Without loss of generality, in this work we adopt Lorentzian hybridization
\begin{equation}
\Gamma_{\alpha m}(\omega)=\frac{\Delta _{\alpha m}W_{\alpha m}^{2}}{\left( \omega -\mu _{\alpha m}\right) ^{2}+W_{\alpha m}^{2}} \label{Lorentzian}
\end{equation}
with $\Delta_{\alpha m}$ being the effective coupling strength between dot $m$ and lead $\alpha m$, and $W_{\alpha m}$ being the width of the conduction band of lead $\alpha m$. In numerical calculations, we adopt $W_{Lt}=0.1$ meV to simulate a strongly energy dependent hybridization $\Gamma_{Lt}(\omega)$, while $W_{Rt}=W_{Lb}=W_{Rb}=4$ meV to simulate constant hybridizations for the other leads. Moreover, we set $\Delta_{Lt}=\Delta_{Rt}=\Delta_{Lb}=\Delta_{Rb}\equiv\Delta$ for simplicity.

\subsection{The TME approach}\label{theTME}
The TME approach is routinely used for transport through interacting quantum dots involving Coulomb interactions \cite{PhysRevB.65.115332,PhysRevB.65.045317,PhysRevB.75.165303,PhysRevB.85.045325}, electron-phonon couplings \cite{PhysRevB.70.195107,PhysRevB.74.205438,cao2017thermoelectric}, spin exchange interactions \cite{PhysRevB.73.235304,PhysRevB.73.235305,PhysRevB.76.054448}, and so on. We summarize the three steps of studying capacitively coupled double quantum dots using the TME.

\textit{Step 1: Setting up the master equation.} The relevant quantum states are $\vert m\rangle=|0\rangle,|t\rangle,|b\rangle,|2\rangle$ with respective energies $E_m=0, \varepsilon_t, \varepsilon_b, \varepsilon_t+\varepsilon_b+U$, representing the empty state, singly occupied state on the top dot, singly occupied state on the bottom dot, and doubly occupied state, respectively. Formally, the master equation describing the time evolution of the probability $P_m$ of the nonequilibrium occupation of state $\vert m\rangle$ is
\begin{equation}
\frac{dP_m(t)}{dt}=-\sum_{n\ne m}\gamma_{m\rightarrow n}P_m+\sum_{n\ne m}\gamma_{n\rightarrow m}P_n \label{ME}
\end{equation}
with the normalization condition $\sum_m P_m=1$. The first (second) summation gives the rate at which the state $\vert m\rangle$ decays (increases), with $\gamma_{m\rightarrow n}$ being the transition rate from state $\vert m\rangle$ to $\vert n\rangle$ induced by electron tunneling. For time-independent systems, one focuses on the steady-state transport where $\frac{dP_m(t)}{dt}=0$. Henceforth, Eq.~(\ref{ME}) can be written in the form $W P=0$, with $W$ being a $4\times 4$ matrix and $P=\{P_0,P_t,P_b,P_2\}^T$.

\begin{figure}[t!]
\centering
\includegraphics[width=\columnwidth]{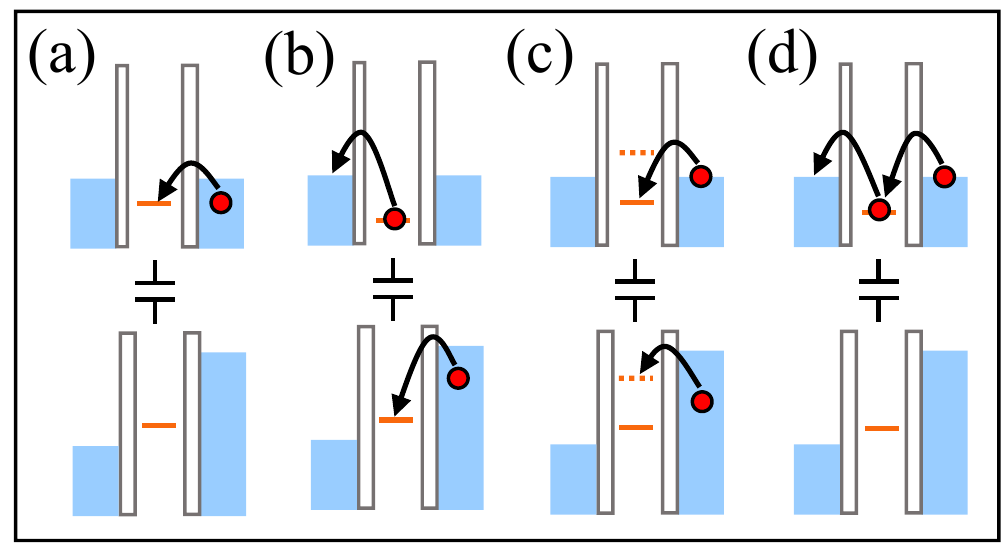}
\caption{Exemplary illustrations of (a) second-order and (b)-(d) fourth-order electron tunneling processes captured by the TME approach. (a) The sequential tunneling process which transfers an electron into the top dot from lead $Rt$. In (b) inelastic cotunneling, (c) pair tunneling, and (d) elastic cotunneling, two electrons are transferred coherently between the dots and leads. The solid (dashed) lines between the tunneling barriers denote the dot level $\varepsilon_{t/b}$ ($\varepsilon_{t/b}+U$). Note that the TME overlooks additional fourth-order tunneling processes without explicit illustrations (see Sec.~\ref{remarks}).}\label{Fig:MEillustration}
\end{figure}

\textit{Step 2: Calculating the transition rate $\gamma_{m\rightarrow n}$.} With the weak-coupling assumption that $\Gamma_{\alpha m}$ is smaller than the other energy scales, the transition rates $\gamma_{m\rightarrow n}$ can be obtained by the generalized Fermi's golden rule \cite{bruus2004many},
\begin{equation}
\gamma_{m\rightarrow n}  = \frac{ 2 \pi}{\hbar} \sum_{i,f} \rho_{i} |\langle f n|T|m i\rangle |^2 \delta (E_{\textrm{final}}-E_{\textrm{initial}}),\label{FGR}
\end{equation}
with $|f n\rangle=|f\rangle \otimes |n\rangle$ and $|m i\rangle=|m\rangle \otimes |i\rangle$. $\vert i\rangle$ ($\vert f\rangle$) denotes the initial (final) lead state with energy $E_{\textrm{leads},i(f)}$. $E_{\textrm{initial}}=E_m+E_{\textrm{leads},i}$ and $E_{\textrm{final}}=E_n+E_{\textrm{leads},f}$ are the initial and final state energies of the entire system, respectively. The probability $\rho_i$ of lead state $\vert i\rangle$ is given by the equilibrium grand-canonical Gibbs distribution.

The $T$-matrix in Eq.~(\ref{FGR}) obeys the recursive relation
\begin{equation}
T=H_T+H_T\frac{1}{E_{\textrm{initial}}-H_0+i\eta}T,\label{TM}
\end{equation}
where $H_0=H_{\textrm{DQD}}+\sum_{\alpha m} H_{\alpha m}$ and $\eta$ is a positive infinitesimal. By truncating the $T$-matrix up to second order in $H_T$, the $W$ matrix mentioned in step 1 becomes $W=\sum_{\alpha=L,R}W_\alpha$, with $W_\alpha$ being
\begin{align}
\left(
\begin{array}
[c]{ccccc}%
(W_\alpha)_{00} &\gamma_{t\rightarrow0}^{Dt\rightarrow\alpha t} &\gamma_{b\rightarrow0}^{Db\rightarrow\alpha b} &\sum_{\beta}\widetilde\gamma_{2\rightarrow0}^{\rightarrow\alpha t,\beta b}\\
\gamma_{0\rightarrow t}^{\alpha t\rightarrow Dt} &(W_\alpha)_{tt} &\sum_{\beta}\widetilde\gamma_{b\rightarrow t}^{\alpha t\rightarrow\beta b} &\gamma_{2\rightarrow t}^{Db\rightarrow\alpha b}\\
\gamma_{0\rightarrow b}^{\alpha b\rightarrow Db} &\sum_{\beta}\widetilde\gamma_{t\rightarrow b}^{\alpha b\rightarrow\beta t} &(W_\alpha)_{bb} &\gamma_{2\rightarrow b}^{Dt\rightarrow\alpha t}\\
\sum_{\beta}\widetilde\gamma_{0\rightarrow2}^{\alpha t,\beta b\rightarrow} &\gamma_{t\rightarrow2}^{\alpha b\rightarrow Db} &\gamma_{b\rightarrow2}^{\alpha t\rightarrow Dt} &(W_\alpha)_{22}
\end{array}
\right),\label{WA}
\end{align}
with the diagonal elements $(W_\alpha)_{ii}=-\sum_{j\ne i}(W_\alpha)_{ji}$. The notation $\gamma\propto t^2_{k\alpha m}$ ($\widetilde\gamma\propto t^4_{k\alpha m}$) denotes the transition rate of the second-order (fourth-order) tunneling process, in which one (two) electron(s) is sequentially (coherently) transferred between the dots and leads. The superscripts indicate the associated electron tunnelings. Concretely, $\gamma_{0\rightarrow t}^{R t\rightarrow Dt}$ describes that an electron in lead $Rt$ tunnels into the top dot, resulting in the transition from state $|0\rangle$ to $|t\rangle$, as illustrated in Fig.~\ref{Fig:MEillustration}(a). $\widetilde\gamma_{t\rightarrow b}^{Rb\rightarrow Lt}$ describes an inelastic cotunneling process in which an electron tunnels from the top dot into lead $Lt$ together with an electron in lead $Rb$ entering the bottom dot coherently [see Fig.~\ref{Fig:MEillustration}(b)]. Similarly, $\widetilde\gamma_{0\rightarrow 2}^{Rt,Rb\rightarrow}$ and $\widetilde\gamma_{t\rightarrow t}^{Rt \rightarrow Lt}$ describe a pair tunneling and elastic cotunneling, respectively, as illustrated in Figs.~\ref{Fig:MEillustration}(c) and \ref{Fig:MEillustration}(d). The detailed transition rates are listed in Appendix \ref{app}.

\textit{Step 3: Calculating the charge currents.} Once the probabilities $P_m$ are obtained by solving the master equation, the charge current flowing out of lead $\alpha m$ is
\begin{equation}
I_{\alpha m}=(-e)[I_{\alpha m}^{\textrm{2nd}}+I_{\alpha m}^{\textrm{4th,ela}}+I_{\alpha m}^{\textrm{4th,inela}}+I_{\alpha m}^{\textrm{4th,pair}}],
\end{equation}
with the second-order and fourth-order tunneling contributions
\begin{eqnarray}
I_{\alpha m}^{\textrm{2nd}}&=&\gamma_{0\rightarrow m}^{\alpha m\rightarrow Dm}  P_{0}-\gamma_{m\rightarrow0}^{Dm\rightarrow\alpha m}P_{m}\notag\\
&&+\gamma_{\bar{m}\rightarrow2}^{\alpha m\rightarrow Dm} P_{\bar{m}}-\gamma_{2\rightarrow\bar{m}}^{Dm\rightarrow\alpha m}P_{2},\label{I2}\\
I_{\alpha m}^{\textrm{4th,ela}}&=&\left(\widetilde \gamma_{m\rightarrow m}^{\alpha m\rightarrow\bar{\alpha}m}-\widetilde\gamma_{m\rightarrow m}^{\bar{\alpha}m\rightarrow\alpha m}\right)  P_{m}\notag\\
&&+\left(\widetilde \gamma_{\bar m\rightarrow \bar m}^{\alpha m\rightarrow\bar{\alpha}m}-\widetilde\gamma_{\bar m\rightarrow \bar m}^{\bar{\alpha}m\rightarrow\alpha m}\right)  P_{\bar m},\label{I4ela}\\
I_{\alpha m}^{\textrm{4th,inela}}&=&-\left(\widetilde\gamma_{m\rightarrow\bar{m}}^{\alpha\bar{m}\rightarrow\alpha m}+\widetilde\gamma_{m\rightarrow\bar{m}}^{\bar{\alpha}\bar{m}\rightarrow\alpha m}\right)  P_{m}\notag \\
&&+\left(\widetilde\gamma_{\bar{m}\rightarrow m}^{\alpha m\rightarrow\alpha\bar{m}}+\widetilde\gamma_{\bar{m}\rightarrow m}^{\alpha m\rightarrow\bar{\alpha}\bar{m}}\right)  P_{\bar{m}},\label{I4inela}\\
I_{\alpha m}^{\textrm{4th,pair}}&=&\left(\widetilde\gamma_{0\rightarrow2}^{\alpha m,\bar{\alpha}\bar{m}\rightarrow}+\widetilde\gamma_{0\rightarrow2}^{\alpha m,\alpha\bar{m}\rightarrow}\right)  P_{0}\notag\\
&&-\left(\widetilde\gamma_{2\rightarrow0}^{\rightarrow \alpha m,\bar{\alpha}\bar{m}}+\widetilde\gamma_{2\rightarrow0}^{\rightarrow\alpha m,\alpha\bar{m}}\right)  P_{2},\label{I4pair}
\end{eqnarray}
where $\bar\alpha$ ($\bar m$) denotes the opposite index of $\alpha$ ($m$). We define the drag and drive currents as $I_{\textrm{drag}}=I_{Lt}$ and $I_{\textrm{drive}}=I_{Lb}$, respectively. The average dot occupancies can be obtained as $\langle N_t \rangle=P_t+P_2$ and $\langle N_b \rangle=P_b+P_2$.

The pioneering work \cite{PhysRevLett.104.076801} addressing the Coulomb drag in double quantum dots by the TME considered only the second-order (sequential) tunneling processes, i.e., setting all $\widetilde\gamma=0$. We refer to this operation as S-TME hereafter. Recent works \cite{PhysRevLett.116.196801,PhysRevLett.117.066602} have uncovered the significance of the fourth-order tunneling processes in understanding the experiments, as we mentioned in Sec.~\ref{int}.

\subsection{The HEOM approach}\label{HEOMappr}
We briefly outline in this section the HEOM formalism \cite{jin2008exact,zheng2009numerical,PhysRevLett.109.266403}, which was developed for treating quantum open systems consisting of a system (quantum dots here), reservoirs (metallic leads here), and system-reservoir couplings (dot-lead couplings here). In the quantum dissipation theory, the quantity of primary interest is the reduced system density matrix $\rho(t) \equiv {\rm tr}_{\rm{res}}[\rho_{\rm{total}}(t)]$, with ${\rm tr}_{\rm{res}}$ the trace over all reservoir degrees of freedom, which can be obtained through
\begin{equation}
\rho(t)=\mathcal{U}(t,t_{0})\rho(t_{0}) \label{evolutionequation}
\end{equation}
with the Liouville-space propagator
\begin{equation}
\mathcal{U}(t,t_0)=\int^{t}_{t_{0}}\mathcal{D}\psi \int^t_{t_0} \mathcal{D}\psi' e^{i\mathcal{S}[\psi]} \mathcal{F}[\psi,\psi'] e^{-i\mathcal{S}[\psi']}.
\end{equation}
Here, $\mathcal{S}[\psi]$ is the classical action functional of the system and $\mathcal{F}[\psi,\psi']$ accounts for the influence of the reservoirs on the properties of the system. The latter has a rather complicated form and is referred to in Ref.~\onlinecite{jin2008exact}. However, we mention that the system-reservoir coupling enters $\mathcal{F}[\psi,\psi']$ exclusively through the reservoir correlation function $C_{\alpha mn}^\sigma(t,\tau)$, with $\alpha$ being the reservoir index, $m$ and $n$ being the system states, and $\sigma=\pm$. For our model,
\begin{equation}
C^\sigma_{\alpha mn}(t,t')=\delta_{mn}\tilde C^\sigma_{\alpha m}(t-t'),
\end{equation}
\begin{equation}
\tilde C^\sigma_{\alpha m}(t)=\int^\infty_{-\infty}d\omega e^{i\sigma\omega t}\Gamma_{\alpha m}(\omega)f_{\alpha}^\sigma(\omega),\label{eqcorr}
\end{equation}
where $\Gamma_{\alpha m}(\omega)$ is defined in Eq.~(\ref{Lorentzian}) and $f_{\alpha}^\sigma(\omega)=1/[1+e^{\sigma(\hbar\omega-\mu_{\alpha})/k_BT_\alpha}]$ is the Fermi distribution for an electron ($\sigma=+$) or hole ($\sigma=-$).

The central step towards establishing a closed HEOM is the decomposition of $\tilde C^\sigma_{\alpha m}(t)$ into a exponential series. Formally, by using the contour integral with the Cauchy residue theorem, Eq.~(\ref{eqcorr}) can be recast as
\begin{equation}
\tilde C^\sigma_{\alpha m}(t)=\sum_{q=1}^\infty\eta^\sigma_{\alpha m q}e^{-\gamma^\sigma_{\alpha m q} t}, \label{decomposition}
\end{equation}
where $\{\gamma^\sigma_{\alpha m q}\}$ are related to the poles of $\Gamma_{\alpha m}(\omega)$ and $f_{\alpha}^\sigma(\omega)$. In practical calculations, we take a truncation of Eq.~(\ref{decomposition}) and retain the $Q$ leading terms. For the Pad\'{e} decomposition of $f_{\alpha}^\sigma(\omega)$ the $Q$ value is determined by a sufficiently small discrepancy between the approximated Fermi distribution and the exact one at a certain temperature. Starting with Eq.~(\ref{evolutionequation}) and the decomposition of $\tilde C^\sigma_{\alpha m}(t)$, the formally exact HEOM without any approximations can be derived as
\begin{align}
   \dot\rho^{(l)}_{j_1\cdots j_l} =& -\Big(i{\cal L} + \sum_{r=1}^l \gamma_{j_r}\Big)\rho^{(l)}_{j_1\cdots j_l}
     -i \sum_{j}\!
     {\cal A}_{\bar j}\, \rho^{(l+1)}_{j_1\cdots j_lj}
\nonumber\\ &
    -i \sum_{r=1}^{l}(-1)^{l-r}\, {\cal C}_{j_r}\,
     \rho^{(l-1)}_{j_1\cdots j_{r-1}j_{r+1}\cdots j_l},\label{HEOM}
\end{align}
where $\rho^{(0)}(t)=\rho(t)$ and $\{\rho^{(l)}_{j_1\cdots j_l}(t); l=1,\cdots,L\}$ are the auxiliary density matrices, which are fermionic (bonsonic) operators for $l$ being odd (even) integers, with $L$ being the truncated tier level. We note that the hierarchy is self-contained at $L=2$ for noninteracting systems, while for systems involving Coulomb interactions the solution to the HEOM must go through systematic tests to confirm its convergence versus $L$. In practice, a relatively low $L$ ($\approx 4$) is usually sufficient to yield quantitatively converged results. The multicomponent index $j \equiv (\sigma\alpha mq)$. The superoperators $\mathcal{L}$, ${\cal A}_{\bar j}\equiv {\cal A}_m^{\bar\sigma}$, and ${\cal C}_{j}\equiv {\cal C}^\sigma_{\alpha mq}$ are defined via their actions on a fermionic/bosonic operator $O$ as $\mathcal{L}O \equiv \hbar^{-1}[H_{\textrm{sys}}, O]_-$, ${\cal A}_m^{\bar \sigma}O\equiv [d_m^{\bar\sigma},O]_\mp$, and ${\cal C}^\sigma_{\alpha mq}O\equiv\eta^\sigma_{\alpha mq}d^\sigma_m O\pm (\eta^{\bar \sigma}_{\alpha mq})^*O d^\sigma_m$, with $d^+_m\equiv d^\dag_m$ and $d^-_m\equiv d_m$. After solving the HEOM, the charge current from lead $\alpha$ to dot $m$ can be obtained with
\begin{equation}
I_{\alpha m}(t)=ie\{\mathrm{tr}_{\textrm{sys}}[{\rho^\dag_{\alpha m}(t)d_{m} -d^\dag_{m}\rho^-_{\alpha m}(t)}]\},
\end{equation}
where $\rho^\dag_{\alpha m}=(\rho^-_{\alpha m})^\dag$ is the first-tier auxiliary density operator. The current conservation is respected within the numerical precision.

The numerical implementation of the HEOM formalism, usually termed HEOM, can capture the combined effects of system-reservoir dissipation, many-body interactions, and non-Markovian memory in a nonperturbative manner. The HEOM approach is applicable to both static and dynamic properties of diverse quantum impurity systems \cite{zheng2008dynamic,zheng2008dynamicElectronic,PhysRevLett.111.086601,PhysRevB.88.035129,PhysRevB.90.165116,cheng2015time,PhysRevB.91.205106,PhysRevB.94.245105,li2017corrected,cheng2018transient,PhysRevB.98.115133}. Moreover, it has also been combined with the density-functional theory to study the correlated electronic structure of adsorbed magnetic molecules \cite{wang2014understanding,PhysRevB.93.125114,wang2016anisotropy}.

The accuracy of the HEOM approach for the single-impurity Anderson model has been explicitly demonstrated. Specifically, the HEOM achieves the same level of accuracy as the full density-matrix numerical renormalization group for the local density of states \cite{PhysRevLett.109.266403}. In addition, the steady-state current obtained with the HEOM reproduces the ones calculated by the real-time quantum Monte Carlo, time-dependent density-matrix renormalization group, functional renormalization group, and iterative summation of real-time path integral approaches \cite{ye2016heom}. In our previous works \cite{PhysRevLett.109.266403,li2017corrected,PhysRevB.98.115133,cheng2018transient,cheng2015time}, we employed the HEOM to study the intricate nonequilibrium Kondo effects in both single and double quantum dots. The extension of our HEOM method to solving the present Coulomb drag problem is straightforward. Therefore, we shall take the converged results obtained with the HEOM as the benchmark to evaluate the performance of the TME.

\subsection{TME versus HEOM}\label{HEOMappr}\label{remarks}
We make some remarks on the TME and HEOM approaches. (i) As the dot-lead tunnel couplings are treated as the perturbative terms, the TME is commonly believed to be reliable if the coupling strengths are smaller than the other energy scales. (ii) As pointed out in Refs.~\onlinecite{PhysRevB.82.235307} and \onlinecite{PhysRevB.82.045316}, the TME is not a systematic perturbation theory. Specifically, the TME includes only the fourth-order tunneling processes with transition rates expressed in the form of a squared matrix element [see Eq.~(\ref{FGR})], rather than all the generic fourth-order tunnelings. (iii) The fourth-order tunneling processes captured by the TME transfer two electrons between the dots and leads, as illustrated in Figs.~\ref{Fig:MEillustration}(b)-\ref{Fig:MEillustration}(d). However, in fact, there also exist fourth-order tunneling processes which merely transfer one electron, while a second electron undergoes a virtual transition with charge fluctuations \cite{PhysRevB.82.235307}, which we refer to as fourth-order SETs. These processes have no explicit illustrations like Fig.~\ref{Fig:MEillustration} and are completely overlooked by the TME \cite{PhysRevB.82.235307}. As we will show in Sec.~\ref{results}, it is the overlooked fourth-order SETs that are responsible for the failure of the TME. (iv) Unlike the TME, the HEOM is a nonperturbative numerical approach which can deal with interacting quantum dot systems to a desired precision without restriction of the relevant energy scales. The main disadvantage of the HEOM approach lies in the increasing computational cost as the temperature decreases and/or the reservoir number increases.

\begin{figure}[t!]
\centering
\includegraphics[width=\columnwidth]{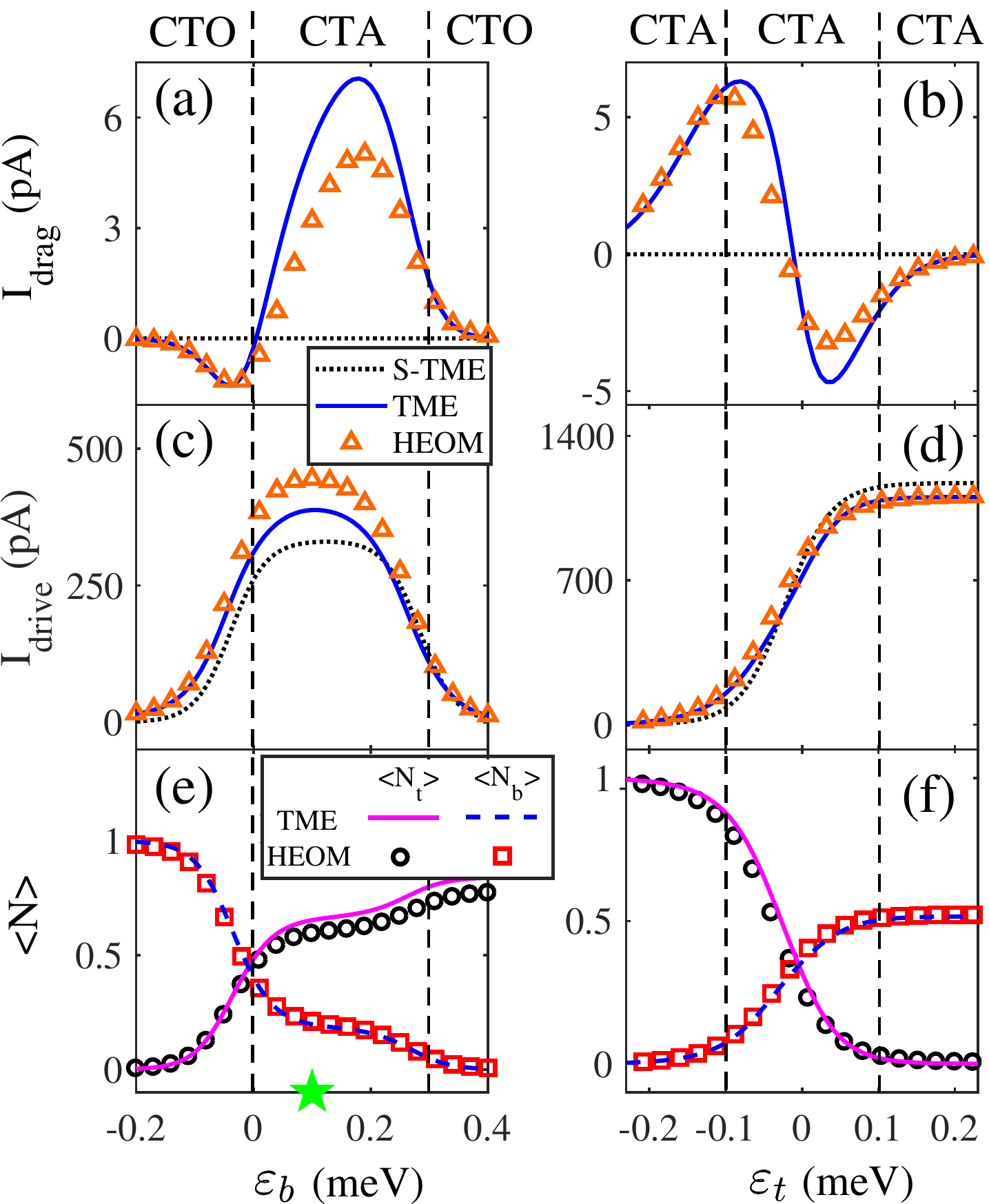}
\caption{Currents and average dot occupancies obtained by different approaches as a function of $\varepsilon_b$ (left panels with $\varepsilon_t=-0.05$ meV) and $\varepsilon_t$ (right panels with $\varepsilon_b=0.1$ meV). As separated by the vertical dashed lines, cotunneling-only (CTO) and cotunneling-assisted (CTA) drag mechanisms dominate at different dot-level configurations. In both columns, the TME fails in the regions dominated by CTA drag, with average dot occupancies largely departing from $0$ and $1$. The green star in (e) marks the dot-level configuration studied in Fig.~\ref{Fig:Temp}. We adopt $\mu_{Rb}=0.3$ meV, $\Delta=0.01$ meV, $k_B T=0.03$ meV, and $U=4$ meV. }\label{Fig:dotlevel}
\end{figure}

\section{Numerical results and discussion}\label{results}
In this section, we compare the results obtained by the S-TME, TME, and HEOM approaches in the weak dot-lead coupling regime. The performance of the TME is quantified in terms of relative deviation of current $|I^{\textrm{TME}}-I^{\textrm{HEOM}}|/|I^{\textrm{HEOM}}|$. In all calculations below, the parameters are chosen within the ranges measured in Coulomb drag experiments \cite{bischoff2015measurement,PhysRevLett.117.066602}. To simplify the analysis, we first consider in Sec.~\ref{LUL} the large-$U$ limit where the doubly occupied state is excluded. Afterwards, the finite-$U$ cases are discussed in Sec.~\ref{FUS}. The small-$U$ case ($U<\Delta$) is not studied as the TME is invalid in this regime.

\subsection{Large $U$ limit}\label{LUL}
We start with $U=4$ meV, much larger than the other relevant energy scales, to exclude the doubly occupied state. This is possible in a graphene based double quantum dot device where $U$ is about $10$ meV \cite{bischoff2015measurement}. In Figs.~\ref{Fig:dotlevel}(a) and \ref{Fig:dotlevel}(b), we compare the drag currents obtained by different approaches as a function of the dot levels $\varepsilon_b$ and $\varepsilon_t$, respectively. Clearly, qualitative current evolutions are well captured by the TME; however, quantitative agreements are only partially achieved. Particularly, considerable relative deviations of the drag current are observed, e.g., $65\%$ for $\varepsilon_b=0.1$ meV in Fig.~\ref{Fig:dotlevel}(a) and $45\%$ for $\varepsilon_t=0.05$ meV in Fig.~\ref{Fig:dotlevel}(b), which are much larger than those of the drive current exhibited in Figs.~\ref{Fig:dotlevel}(c) and \ref{Fig:dotlevel}(d). Such a failure of the TME for Coulomb drag at the quantitative level has not been realized previously.

\begin{figure}[t!]
\centering
\includegraphics[width=\columnwidth]{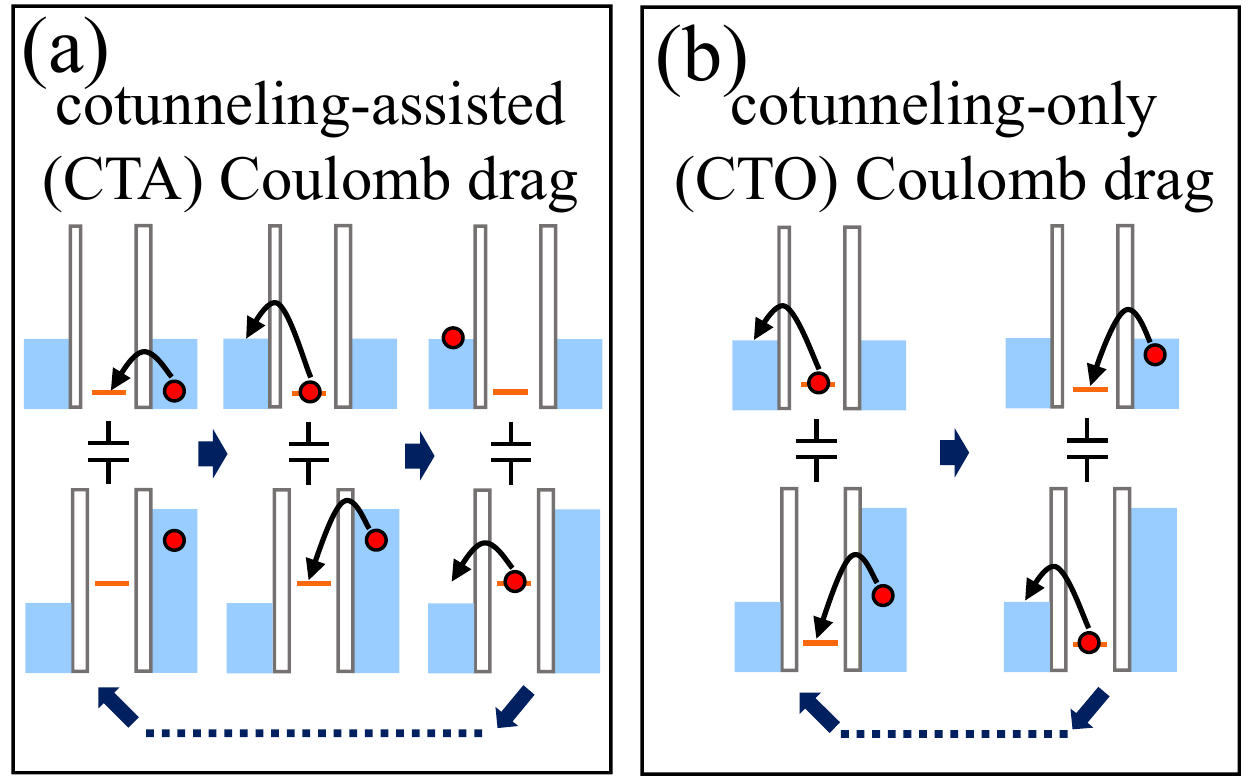}
\caption{Illustrations of (a) cotunneling-assisted (CTA) and (b) cotunneling-only (CTO) Coulomb drag mechanisms for $\varepsilon_t$ below the Fermi level. (a) For $\varepsilon_b$ inside the bias window, two sequential tunnelings are assisted by an intermediate nonlocal cotunneling to drag an electron across the top dot. (b) For $\varepsilon_b$ outside the bias window, sequential tunnelings are energetically prohibited; however, two successive nonlocal cotunneling processes can combine to drag an electron. Note that CTA can also work for $\varepsilon_t$ above the Fermi level.}\label{Fig:illus}
\end{figure}

The above drag currents are dominated by different drag mechanisms, depending on the dot-level configuration. As illustrated in Fig.~\ref{Fig:illus}(a), for $\varepsilon_t$ below the Fermi level and $\varepsilon_b$ inside the bias window, two sequential tunneling processes are assisted by an intermediate nonlocal cotunneling process to drag an electron from lead $Rt$ to $Lt$. By contrast, in Fig.~\ref{Fig:illus}(b), for $\varepsilon_b$ outside the bias window, sequential tunnelings are energetically prohibited; however, two successive nonlocal cotunneling processes can combine to drag an electron across the top dot. These two mechanisms are known as cotunneling-assisted and cotunneling-only Coulomb drag \cite{PhysRevLett.116.196801}, respectively. Likewise, electrons can also be dragged from lead $Lt$ to $Rt$ simultaneously. A directional $I_{\textrm{drag}}$ is generated if the drag currents in the two directions have different amplitudes, which is realized under asymmetric and energy-dependent hybridizations in the drag circuit. In Fig.~\ref{Fig:dotlevel}, the regions dominated by different drag mechanisms are indicated and separated by vertical dashed lines.

We explain the failure of the TME as follows. In Figs.~\ref{Fig:dotlevel}(a) and \ref{Fig:dotlevel}(b), the drag currents obtained by the S-TME are pinned at zero, consistent with previous works \cite{PhysRevLett.116.196801,PhysRevLett.117.066602} in which only sequential tunneling does not manage to induce a drag current without the doubly occupied state. The underlying physics will be clear in the next section. The comparisons between the drag currents obtained by the S-TME and TME indicate that the fourth-order tunneling processes make a leading-order correction to the drag current. As remarked in Sec.~\ref{remarks}, the TME can capture only part of the fourth-order tunneling processes. This necessarily leads to a striking inaccuracy if the overlooked fourth-order tunneling processes make a leading-order contribution to the drag current comparable to the other fourth-order tunnelings captured by the TME. As shown in both the left and right columns in Fig.~\ref{Fig:dotlevel}, the TME fails in the regions between the two vertical dashed lines where cotunneling-assisted drag dominates. The associated average dot occupancies shown in Figs.~\ref{Fig:dotlevel}(e) and \ref{Fig:dotlevel}(f) largely depart from $0$ and $1$, indicating significant charge fluctuations on dots. In view of this, we attribute the failure of the TME to the fact that the fourth-order SETs possessing an intermediate charge fluctuation of the initial or final states are completely overlooked by the TME \cite{PhysRevB.82.235307}. In the regions where the TME works well, the charge fluctuations on dots are weak, rendering a quantitatively satisfying drag current even though the fourth-order SETs are overlooked.

Different from the drag currents, the drive currents obtained by the S-TME are nonzero, as shown in Figs.~\ref{Fig:dotlevel}(c) and \ref{Fig:dotlevel}(d). This is because the temperature-induced charge fluctuation on the drag dot diminishes the Coulomb blockade on the drive dot, such that a drive current can be induced by the bias voltage. The comparisons between the drive currents obtained with the S-TME and TME indicate that the fourth-order tunneling processes give only a next-to-leading-order correction to the drive current. As a result, for any dot-level configuration the drive current obtained by the TME is far more accurate than the drag current.

\begin{figure}[t!]
\centering
\includegraphics[width=\columnwidth]{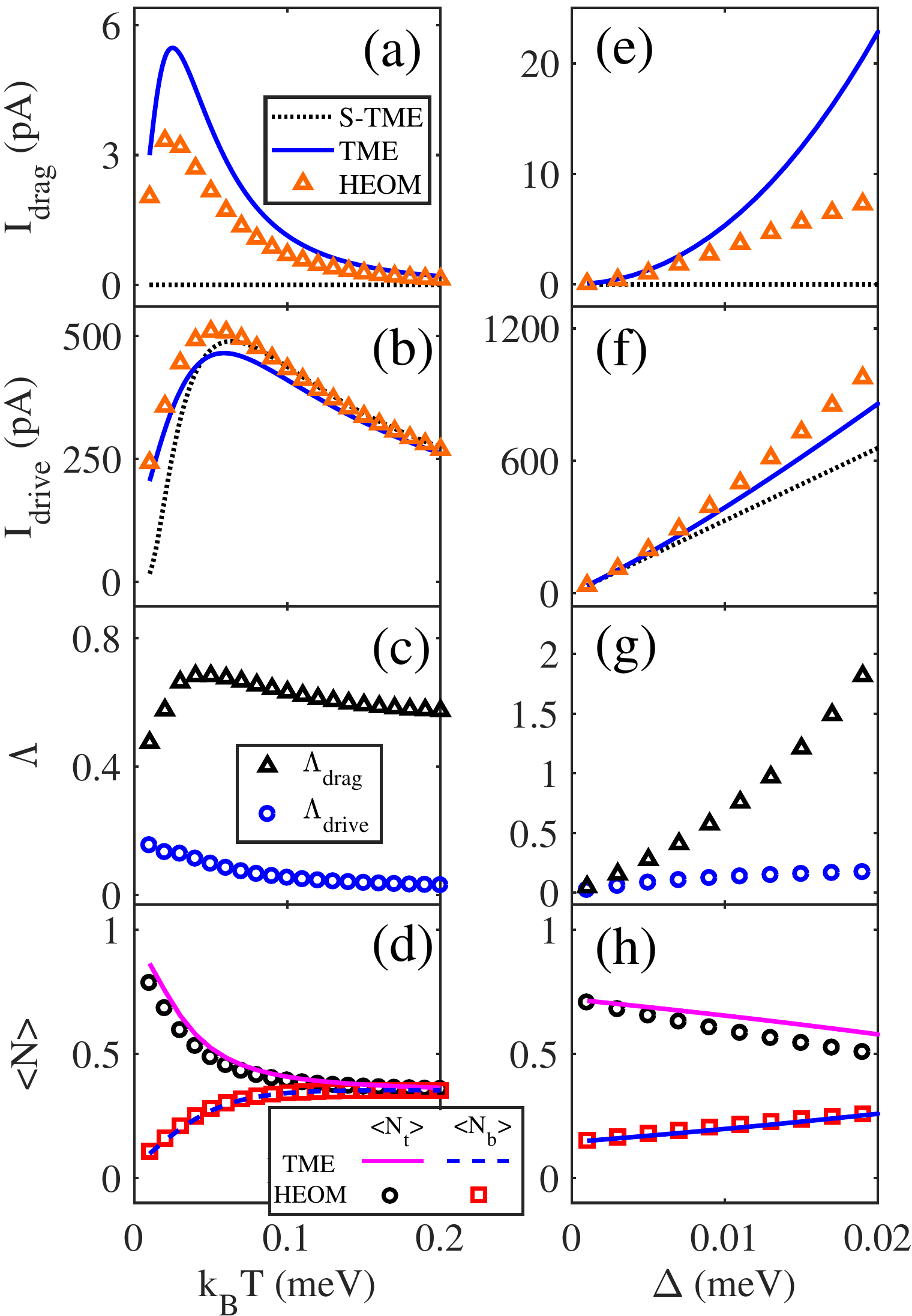}
\caption{Currents and average dot occupancies obtained with different approaches as a function of temperature (left panels with $\Delta=0.01$ meV) and effective coupling strength (right panels with $k_B T=0.03$ meV) at the dot-level configuration marked by the green star in Fig.~\ref{Fig:dotlevel}(c) where cotunneling-assisted drag dominates. The relative deviations of current $\Lambda=|I^{\textrm{TME}}-I^{\textrm{HEOM}}|/|I^{\textrm{HEOM}}|$ are shown in (c) and (g), which further confirm the failure of the TME in obtaining a quantitatively satisfying drag current.}\label{Fig:Temp}
\end{figure}

We proceed to confirm the failure of the TME in the cotunneling-assisted-drag-dominated region. In Fig.~\ref{Fig:Temp}, we present the effects of temperature and effective coupling strength on the currents, average dot occupancies, and the relative deviations at the dot-level configuration marked by the green star in Fig.~\ref{Fig:dotlevel}(e). As clearly shown in Figs.~\ref{Fig:Temp}(c) and \ref{Fig:Temp}(g), the relative deviations of the drag current are rather remarkable and much larger than that of the drive current. We note that these results are obtained in the weak-coupling regime where $\Delta<k_BT$. The average dot occupancies in Figs.~\ref{Fig:Temp}(d) and \ref{Fig:Temp}(h) depart more from $0$ and $1$ with the increase of temperature or effective coupling strength, indicating enhanced charge fluctuations on dots.

In Figs.~\ref{Fig:Temp}(a) and \ref{Fig:Temp}(b), both the drag and drive currents exhibit a nonmonotonic temperature dependence but for different reasons. As the temperature increases, thermal energy facilitates the cotunneling process, [see, e.g., the middle of Fig.~\ref{Fig:illus}(a)] since the electron in the top dot can transit to the lead states below the Fermi level, resulting in an increased drag current with enhanced charge fluctuations [Fig.~\ref{Fig:Temp}(d)]. Upon further increasing the temperature, unoccupied lead states with energy $\varepsilon_t$ are available; therefore, sequential tunnelings eventually take over such that electrons shuttle forward and back across the top dot but yield vanishingly small net current. Notice that, although the absolute deviation between the drag currents obtained by the TME and HEOM is reduced with the increase of temperature [Fig.~\ref{Fig:Temp}(a)], the associated relative deviation remains large with a slow decay [black triangles in Fig.~\ref{Fig:Temp}(c)]. As for the nonmonotonic temperature dependence of the drive current, on the one hand, the increased thermal energy enhances the charge fluctuation on the drag dot to diminish the Coulomb blockade on the drive circuit, which leads to an increased drive current. On the other hand, at higher temperatures, the broadened Fermi distributions reduce the occupied (unoccupied) states below (above) the Fermi level of the source (drain) lead, leading to a suppressed drive current.

In Figs.~\ref{Fig:Temp}(e) and \ref{Fig:Temp}(f), as the effective coupling strength is enhanced, the drag and drive currents increase monotonically since the electron tunnelings between dots and leads become much easier. Meanwhile, the deviation between the currents obtained by the TME and HEOM increases because the TME basically works better for a weaker effective coupling strength. We note again that the HEOM is a numerically exact nonperturbative approach requiring no restriction on the relevant energy scales.

\begin{figure}[t!]
\centering
\includegraphics[width=0.9\columnwidth]{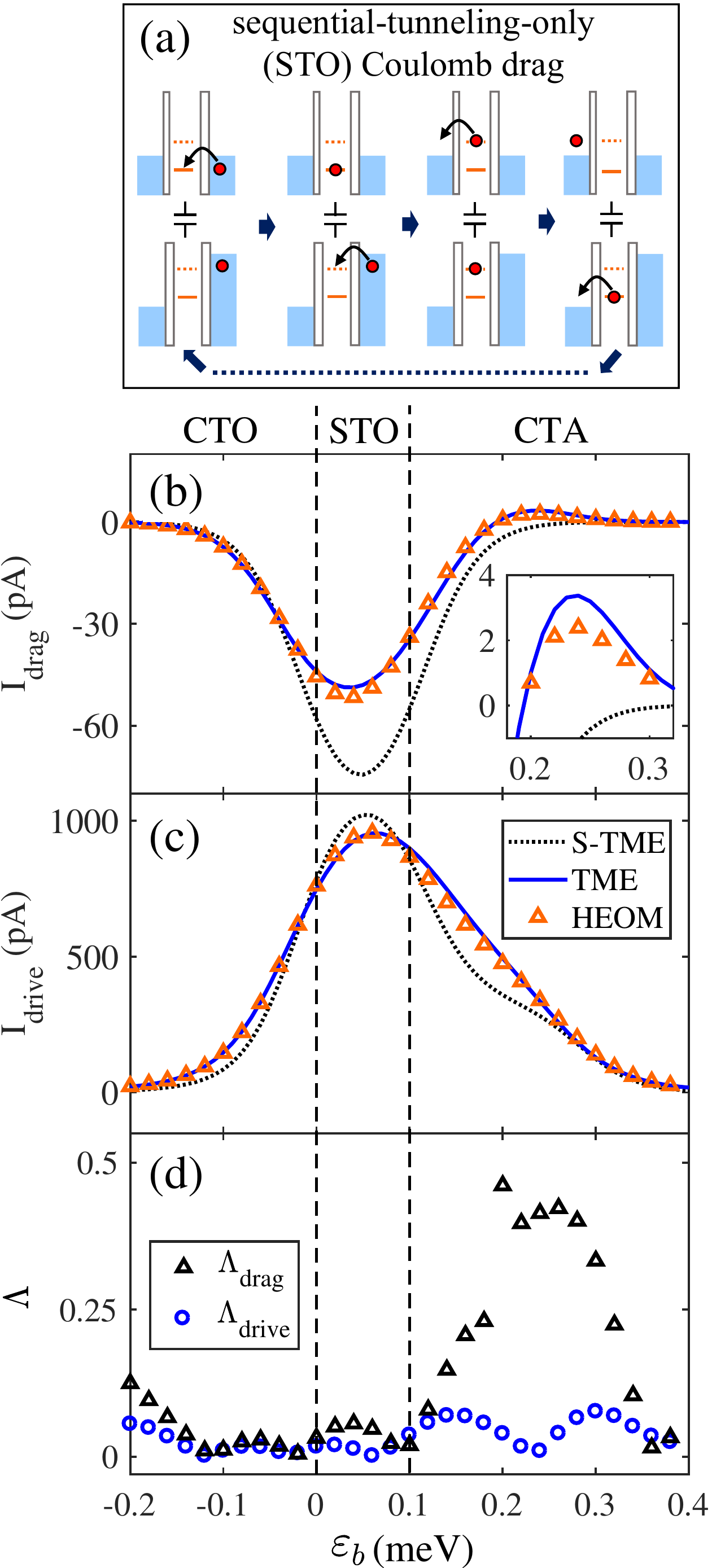}
\caption{(a) Illustration of sequential-tunneling-only drag mechanism. The solid (dashed) lines between the tunneling barriers denote the dot level $\varepsilon_{t/b}$ ($\varepsilon_{t/b}+U$). (b) and (c) Comparisons of the drag and drive currents versus $\varepsilon_b$ obtained with different approaches. The inset in (b) is a zoom-in view for $0.2$ meV $<\varepsilon_b<0.3$ meV. (d) Relative deviation of current $|I^{\textrm{TME}}-I^{\textrm{HEOM}}|/|I^{\textrm{HEOM}}|$. As indicated at the top of (b), three Coulomb drag mechanisms dominate at different regions. The TME fails in the CTA-dominated region, in line with the infinite-$U$ case. The parameters are the same as those in the left column of Fig.~\ref{Fig:dotlevel} except that $U=0.2$ meV.}\label{Fig:finiteU}
\end{figure}

\subsection{Finite-$U$ cases}\label{FUS}
Now we turn to the finite-$U$ cases where the doubly occupied state is permitted such that another drag mechanism can be activated. As illustrated in Fig.~\ref{Fig:finiteU}(a), four successive sequential tunneling processes are combined to drag an electron from lead $Rt$ to $Lt$. This sequential-tunneling-only drag was captured by the S-TME approach \cite{PhysRevLett.104.076801}. It is straightforward from the illustration that this mechanism works in the region (ignoring the thermal broadening of the Fermi distributions)
\begin{eqnarray}
&&\varepsilon_t<\mu_t<\varepsilon_t+U,\label{cond1}\\
&&\mu_{Lb}<\varepsilon_b<\varepsilon_b+U<\mu_{Rb}\label{cond2}.
\end{eqnarray}
In the large-$U$ limit, this drag mechanism is highly suppressed because of the absence of the doubly occupied state required in step 3. This explains the universal zero drag currents obtained by the S-TME in Sec.~\ref{LUL}.

In Fig.~\ref{Fig:finiteU}(b), sequential-tunneling-only drag dominates when $0<\varepsilon_b<0.1$ meV [see Eq.~(\ref{cond2})]. The associated net drag current results from the offset between the drag currents in the two directions, and the profile is proportional to \cite{PhysRevLett.104.076801}
\begin{equation}
(-e)\{f_{t}(\varepsilon_t)\Gamma_{Lt}(\varepsilon_t)[1-f_{t}(\varepsilon_t+U)]\Gamma_{Rt}(\varepsilon_t+U)-L\leftrightarrow R\},\label{sequentialcurrent}
\end{equation}
where $f_{t}(\varepsilon)$ is the Fermi distribution of the top leads. As indicated on the top of Fig.~\ref{Fig:finiteU}(b), cotunneling-only and cotunneling-assisted drag dominate for $-0.2$ meV$<\varepsilon_b<0$ and $\varepsilon_b>0.3$ meV, respectively. As shown in the inset of Fig.~\ref{Fig:finiteU}(b) and more clearly in Fig.~\ref{Fig:finiteU}(d), the TME fails in the cotunneling-assisted-drag-dominated region with considerable relative deviation of the drag current, in line with the large-$U$ limit discussed in the preceding section.

\begin{figure}[t!]
\centering
\includegraphics[width=0.9\columnwidth]{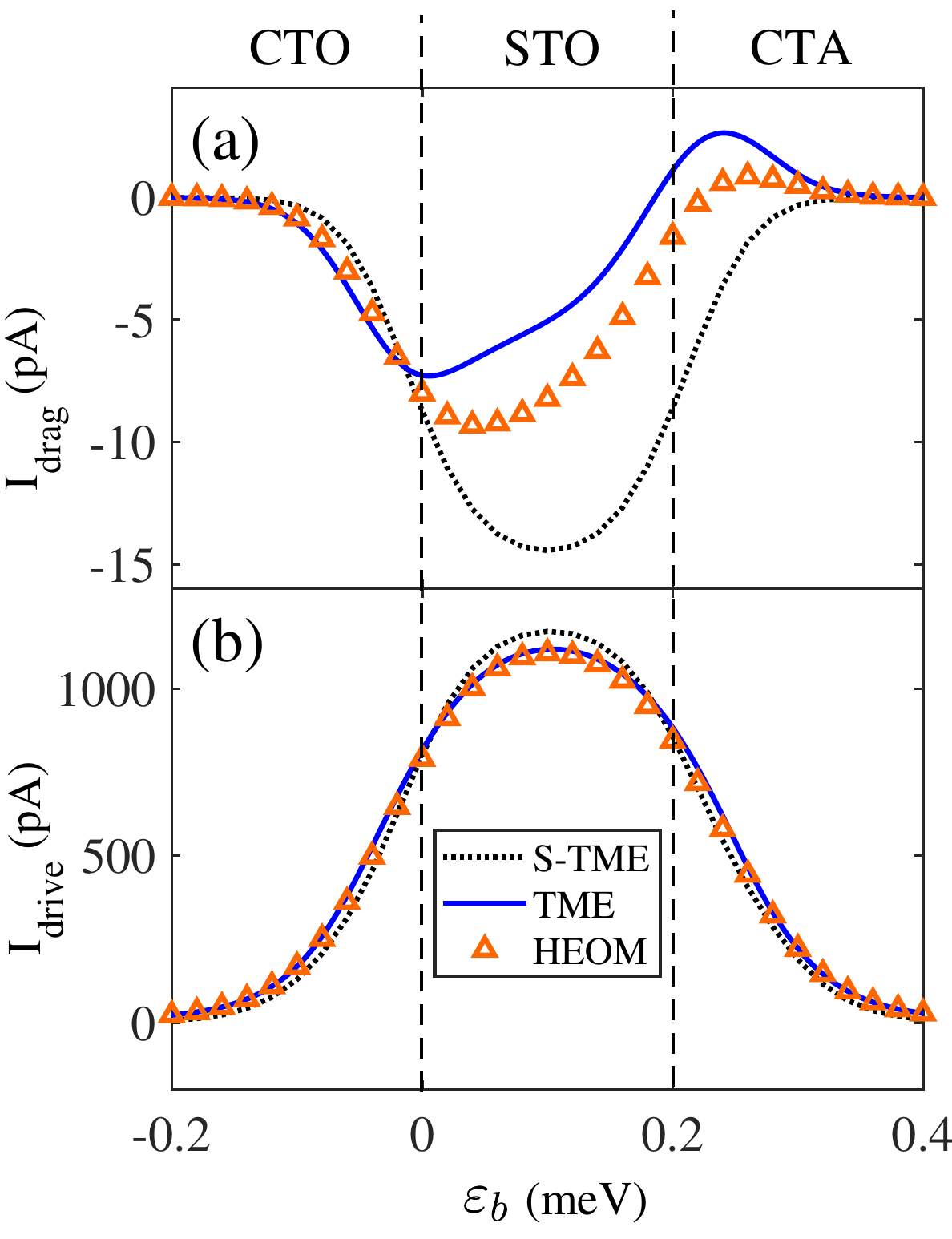}
\caption{Comparisons of the drag and drive currents versus $\varepsilon_b$ obtained with different approaches for $\varepsilon_t$ near $-U/2$. Compared to Fig.~\ref{Fig:finiteU}, the TME fails not only in the CTA-drag-dominated region but also in the STO-drag-dominated region. The parameters are the same as those in Fig.~\ref{Fig:finiteU} except that $U=0.1$ meV and $\varepsilon_t=-0.04$ meV.}\label{Fig:symmetric}
\end{figure}

In the sequential-tunneling-only-drag-dominated region, the drag current is typically one order of magnitude larger than those in the other regions. Meanwhile, the relative deviation between the drag currents obtained by the S-TME and TME is small, implying that the fourth-order tunneling processes give only a next-to-leading-order correction to the drag current. Consequently, even though the TME overlooks the fourth-order SET processes, the resulting drag current almost reproduces the one obtained by the HEOM, as shown in Figs.~\ref{Fig:finiteU}(b) and \ref{Fig:finiteU}(d). However, considerable relative deviations of the drag current show up in the cotunneling-assisted-drag-dominated region, as we mentioned above. When $\varepsilon_b$ is inside the bias window, cotunneling-assisted drag also dominates when $\varepsilon_t<\varepsilon_t+U<0$ or $0<\varepsilon_t<\varepsilon_t+U$, in which case the relative deviation of the drag current would be large if the charge fluctuations on dots are striking. We note that the relative deviation of the drive current is small for the whole range of $\varepsilon_b$, as shown in Figs.~\ref{Fig:finiteU}(c) and \ref{Fig:finiteU}(d), due to the fact that the fourth-order SETs always make a next-to-lading-order correction to the drive current.

Notably, a large relative deviation of drag current may also occur when sequential-tunneling-only drag is dominant, as shown in Fig.~\ref{Fig:symmetric}(a). For $\varepsilon_t$ near $-U/2$, the net drag current obtained by the S-TME is small but nonzero, which is qualitatively consistent with Eq.~(\ref{sequentialcurrent}) by noting that the hybridizations $\Gamma_{Lt}(\varepsilon)$ and $\Gamma_{Rt}(\varepsilon)$ are symmetric about the Fermi level of the top circuit. In this case, the deviations between the drag currents obtained by the S-TME and TME are rather prominent, indicating the non-negligible role of the fourth-order tunneling processes. Moreover, as implied by Fig.~\ref{Fig:finiteU}(a), the successive electron tunneling into and out of the dots would cause large charge fluctuations on dots. As a result, the overlook of the fourth-order SETs possessing charge fluctuations on dots is responsible for the failure of the TME in the sequential-tunneling-only-drag-dominated region in Fig.~\ref{Fig:symmetric}(a). As shown in Fig.~\ref{Fig:symmetric}(b), the relative deviations between the drive currents obtained by the TME and HEOM approaches are negligible, as explained before.

\section{Summary and outlook}\label{Summary}
In summary, with the help of the numerically exact HEOM approach, we have evaluated the performance of the TME for Coulomb drag in capacitively coupled double quantum dots in the weak-coupling regime. It was demonstrated that the TME can capture qualitative current evolutions versus various tunable parameters but only partially succeeds at the quantitative level. Specifically, the TME generally gives a highly inaccurate drag current when large charge fluctuations on dots exist and the fourth-order tunneling processes make a leading-order contribution. This failure of the TME is attributed to its intrinsic deficiency of overlooking the fourth-order SETs possessing intermediate charge fluctuations on dots. Our work suggests that further quantitative studies on Coulomb drag by the TME should be wary of the unreliable regions, which can be identified by comparing the currents and average dot occupations obtained with the S-TME and TME approaches. Actually, obtaining an accurate drag current is crucial for both theoretical and practical interests, such as harvesting a large drag current with the drive voltage being as small as possible. This is promising through designing the circuits in the cotunneling-assisted-drag-dominated region; however, the TME is incompetent for this assignment because of its partial failure in this region.

We note that attempts to use the nonequilibrium Green's function formalism under random phase approximation \cite{moldoveanu2009coulomb,PhysRevB.77.195302}, noncrossing approximation \cite{PhysRevB.96.115414}, and single-bubble approximation \cite{PhysRevB.75.045309,PhysRevB.100.081404} have been made to analyze the Coulomb drag in quantum dots. These approximations can reach some parameter regimes beyond the TME under proper conditions. By contrast, the numerical HEOM approach we employed is capable of exploring Coulomb drag in the whole parameter space, especially the strongly correlated orbital Kondo regime \cite{PhysRevLett.110.046604,PhysRevB.88.235427,bao2014orbital}, in a unified manner. Furthermore, the flexible HEOM approach facilitates studying the influence of additional realistic and tunable ingredients, such as multiple dot levels, interdot Coulomb interaction, alternating voltage, and microwave field, on the Coulomb drag in quantum dot systems.

\section{Acknowledgments}\label{Acknowledgements}
We would like to thank D. S\'{a}nchez, K. Kaasbjerg, and A. J. Keller for their correspondence. This work was financially supported by the China Postdoctoral Science Foundation (Grant No.~2019M651635) and the National Natural Science Foundation of China (Grants No.~11674139, No.~11834005, No.~11604138, No.~61474018, and No.~11504017).

\appendix

\section{Transition rates in the TME approach}\label{app}
For clarity, we present the detailed expressions of the transition rates shown in Sec.~\ref{theTME}. Straightforwardly, by substituting Eqs.~(\ref{HT}) and (\ref{TM}) into Eq.~(\ref{FGR}) with different initial and final quantum dot states, one can obtain the associated transition rates. To be precise, the transition rates of the sequential tunneling processes are
\begin{eqnarray}
\gamma_{0\rightarrow m}^{\alpha m\rightarrow Dm}&&=\frac{1}{\hbar}\Gamma_{\alpha m}\left(  \varepsilon_{m}\right)  f_{\alpha m}\left(  \varepsilon_{m}\right),\\
\gamma_{m\rightarrow2}^{\alpha\bar{m}\rightarrow D\bar{m}}&&=\frac{1}{\hbar}\Gamma_{\alpha\bar{m}}\left(  \varepsilon_{\bar{m}}+U\right)  f_{\alpha\bar{m}}\left(  \varepsilon_{\bar{m}}+U\right),\\
\gamma_{m\rightarrow0}^{Dm\rightarrow\alpha m}&&=\frac{1}{\hbar}\Gamma_{\alpha m}\left(  \varepsilon_{m}\right)  \bar f_{\alpha m}\left(  \varepsilon_{m}\right),\\
\gamma_{2\rightarrow m}^{D\bar{m}\rightarrow\alpha\bar{m}}&&=\frac{1}{\hbar}\Gamma_{\alpha\bar{m}}\left(  \varepsilon_{\bar{m}}+U\right)  \bar f_{\alpha\bar{m}}\left(  \varepsilon_{\bar{m}}+U\right),
\end{eqnarray}
where $f_{\alpha m}\left(\varepsilon\right)  =\frac{1}{\textrm{exp}[\left(  \varepsilon-\mu_{\alpha m}\right)/k_{B}T]+1}$ is the Fermi distribution function of lead $\alpha m$ and $\bar f_{\alpha m}\left(\varepsilon\right)=1-f_{\alpha m}\left(\varepsilon\right)$. Additionally, the notation $f_{-\alpha m}\left(\varepsilon\right)  =\frac{1}{\textrm{exp}[\left(  \varepsilon+\mu_{\alpha m}\right)/k_{B}T]+1}$ appears below.

The transition rates of the fourth-order tunneling processes are
\begin{eqnarray}
&&\widetilde\gamma_{m\rightarrow\bar{m}}^{\alpha\bar{m}\rightarrow\beta m}\notag\\
=&&\int\frac{d\varepsilon}{2\pi\hbar}\Gamma_{\alpha\bar{m}}\left(  \varepsilon\right)  \Gamma_{\beta m}\left(  \varepsilon+\varepsilon_{m}-\varepsilon_{\bar{m}}\right) \Bigl|\frac{1}{\varepsilon-\varepsilon_{\bar{m}}+i\eta}\notag\\
&&-\frac{1}{\varepsilon-\varepsilon_{\bar{m}}-U+i\eta}\Bigr|^{2} f_{\alpha\bar{m}}\left(  \varepsilon\right)   \bar f_{\beta m}\left(\varepsilon+\varepsilon_{m}-\varepsilon_{\bar{m}}\right)\notag\\
=&&A_{\alpha \bar m,\beta m}\left(\varepsilon_{m}-\varepsilon_{\bar{m}},\varepsilon_{\bar{m}}\right)  +A_{\alpha \bar m,\beta m}\left(\varepsilon_{m}-\varepsilon_{\bar{m}},\varepsilon_{\bar{m}}+U\right)\notag\\
&&-B_{\alpha \bar m,\beta m}\left(\varepsilon_{m}-\varepsilon_{\bar{m}},\varepsilon_{\bar{m}},\varepsilon_{\bar{m}}+U\right)\label{A5}
\end{eqnarray}
for the inelastic cotunneling processes,
\begin{eqnarray}
&&\widetilde\gamma_{0\rightarrow2}^{\alpha\bar{m},\beta m\rightarrow}\notag\\
=&&\int\frac{d\varepsilon}{2\pi\hbar}\Gamma_{\alpha\bar{m}}\left(  \varepsilon\right)\Gamma_{\beta m}\left(  E_{2}-\varepsilon\right)\Bigl| \frac{1}{\varepsilon-\varepsilon_{\bar{m}}+i\eta}\notag\\
&&-\frac{1}{\varepsilon-\varepsilon_{\bar{m}}-U+i\eta}\Bigr| ^{2}f_{\alpha\bar{m}}\left(\varepsilon\right)  f_{\beta m}\left(  E_{2}-\varepsilon\right)\notag\\
=&&A_{\alpha \bar m,-\beta m}\left(-E_{2},\varepsilon_{\bar{m}}\right)  +A_{\alpha \bar m,-\beta m}\left(-E_{2},\varepsilon_{\bar{m}}+U\right)\notag\\
&&-B_{\alpha \bar m,-\beta m}\left( -E_{2},\varepsilon_{\bar{m}},\varepsilon_{\bar{m}}+U\right),
\end{eqnarray}
\begin{eqnarray}
&&\widetilde\gamma_{2\rightarrow0}^{\rightarrow\alpha\bar{m},\beta m}\notag\\
=&&\int\frac{d\varepsilon}{2\pi\hbar}\Gamma_{\alpha\bar{m}}\left(  \varepsilon\right)\Gamma_{\beta m}\left(  E_{2}-\varepsilon\right)\Bigl| \frac{1}{\varepsilon-\varepsilon_{\bar{m}}+i\eta}\notag\\
&&-\frac{1}{\varepsilon-\varepsilon_{\bar{m}}-U+i\eta}\Bigr| ^{2} \bar f_{\alpha\bar{m}}\left(  \varepsilon\right)  \bar f_{\beta m}\left(E_{2}-\varepsilon\right)\notag\\
=&&A_{-\beta m,\alpha\bar{m}}\left(E_{2},-\varepsilon_{m}-U\right)  +A_{-\beta m,}\left(E_{2},-\varepsilon_{m}\right)\notag\\
&&-B_{-\beta m,\alpha\bar{m}}\left(E_{2},-\varepsilon_{m}-U,-\varepsilon_{m}\right)
\end{eqnarray}
for the pair tunneling processes, and
\begin{eqnarray}
\widetilde\gamma_{0\rightarrow0}^{\alpha m\rightarrow\bar{\alpha}m}&&=\int\frac{d\varepsilon}{2\pi\hbar}\Gamma_{\alpha m}\left(  \varepsilon\right)\Gamma_{\bar{\alpha}m}\left(  \varepsilon\right) \Bigl| \frac{1}{\varepsilon-\varepsilon_{m}+i\eta}\Bigr|^{2}\notag\\
&& \times f_{\alpha m}\left(\varepsilon\right)  \bar f_{\bar{\alpha}m}\left(  \varepsilon\right)=A_{\alpha m,\bar{\alpha}m}\left(  0,\varepsilon_{m}\right),
\end{eqnarray}
\begin{eqnarray}
\widetilde\gamma_{m\rightarrow m}^{\alpha\bar{m}\rightarrow\bar{\alpha}\bar{m}}&&=\int\frac{d\varepsilon}{2\pi\hbar}\Gamma_{\alpha\bar{m}}\left(\varepsilon\right)  \Gamma_{\bar{\alpha}\bar{m}}\left(  \varepsilon\right)}\Bigl| \frac{1}{\varepsilon-\varepsilon_{\bar{m}}-U+i\eta}\Bigr| ^{2\notag\\
&&\times f_{\alpha\bar{m}}\left(  \varepsilon\right)  \bar f_{\bar{\alpha}\bar{m}}\left(  \varepsilon\right)=A_{\alpha\bar{m},\bar{\alpha}\bar{m}}\left(  0,\varepsilon_{\bar{m}}+U\right),
\end{eqnarray}
\begin{eqnarray}
\widetilde\gamma_{m\rightarrow m}^{\alpha m\rightarrow\bar{\alpha}m}&&=\int\frac{d\varepsilon}{2\pi\hbar}\Gamma_{\alpha m}\left(  \varepsilon\right)\Gamma_{\bar{\alpha}m}\left(  \varepsilon\right) \Bigl| \frac{1}{\varepsilon-\varepsilon_{m}+i\eta}\Bigr| ^{2}\notag\\
&&\times f_{\alpha m}\left(\varepsilon\right)  \bar f_{\bar{\alpha}m}\left(  \varepsilon\right)=A_{\alpha m,\bar{\alpha}m}\left(  0,\varepsilon_{m}\right),
\end{eqnarray}
\begin{eqnarray}
\widetilde\gamma_{2\rightarrow2}^{\alpha m\rightarrow\bar{\alpha}m}&&=\int\frac{d\varepsilon}{2\pi\hbar}\Gamma_{\alpha m}\left(  \varepsilon\right)\Gamma_{\bar{\alpha}m}\left(  \varepsilon\right) \Bigl| \frac{1}{\varepsilon-\varepsilon_{m}-U+i\eta}\Bigr| ^{2}\notag\\
&& \times f_{\alpha m}\left(\varepsilon\right)  \bar f_{\bar{\alpha}m}\left(  \varepsilon\right)=A_{\alpha m,\bar{\alpha}m}\left(  0,\varepsilon_{m}+U\right)\notag\\
\label{A11}
\end{eqnarray}
for the elastic cotunneling processes. The auxiliary functions $A_{\alpha m,\beta n}\left(  r,s\right)$ and $B_{\alpha m,\beta n}\left(  r,s_{1},s_{2}\right)$ in Eqs.~(\ref{A5})-(\ref{A11}) are
\begin{eqnarray}
A_{\alpha m,\beta n}\left(  r,s\right)=&&\int\frac{d\varepsilon}{2\pi\hbar}\Gamma_{\alpha m}\left(\varepsilon\right)  \Gamma_{\beta n}\left(  \varepsilon+r\right) \Bigl|\frac{1}{\varepsilon-s+i\eta}\Bigr| ^{2}\notag\\
&&\times f_{\alpha m}\left(\varepsilon\right)  \bar f_{\beta n}\left(  \varepsilon+r\right) ,\label{funcA}
\end{eqnarray}
\begin{eqnarray}
&&B_{\alpha m,\beta n}\left(  r,s_{1},s_{2}\right)\notag\\
=&&\int\frac{d\varepsilon}{\pi\hbar}\Gamma_{\alpha m}\left(  \varepsilon\right)  \Gamma_{\beta n}\left(  \varepsilon+r\right) \operatorname{Re}\frac{1}{(\varepsilon-s_{1}+i\eta)(\varepsilon-s_{2}-i\eta)}\notag\\
&&\times f_{\alpha m}\left(  \varepsilon\right) \bar f_{\beta n}\left(  \varepsilon+r\right), s_1\ne s_2.\label{funcB}
\end{eqnarray}
The next step is to calculate $A_{\alpha m,\beta n}\left(  r,s\right)$ and $B_{\alpha m,\beta n}\left(  r,s_{1},s_{2}\right)$ using a few tricks. First, we transform Eqs.~(\ref{funcA}) and (\ref{funcB}) to
\begin{eqnarray}
A_{\alpha m,\beta n}\left(  r,s\right)=&&-\frac{1}{2\pi\hbar}n_{B}\left(  \mu_{\beta n}-\mu_{\alpha m}-r\right)\notag\\
&&\times \operatorname{Im} \widetilde A_{\alpha m,\beta n}\left(  r,s\right),\label{transformedA}
\end{eqnarray}
\begin{eqnarray}
B_{\alpha m,\beta n}\left(  r,s_{1},s_{2}\right)=&&-\frac{1}{\pi\hbar}n_{B}\left(  \mu_{\beta n}-\mu_{\alpha m}-r\right)\notag\\
&&\times \operatorname{Im}\widetilde B_{\alpha m,\beta n}\left(r, s_1,s_2\right), \label{transformedB}
\end{eqnarray}
with $n_B(x)=\frac{1}{\textrm{exp}[x/k_BT-1]}$ being the Bose distribution and
\begin{eqnarray}
\widetilde A_{\alpha m,\beta n}\left(  r,s\right)=&& \int\frac{d\varepsilon}{\pi}\frac{1}{\left(  \varepsilon-s\right)  ^{2}+\eta^{2}}\Gamma_{\alpha m}\left(\varepsilon\right) \Gamma_{\beta n}\left(  \varepsilon+r\right)\notag\\
&&\times  \Bigl[  \Psi_{\beta n}^{+}\left(  \varepsilon+r\right)-\Psi_{\alpha m}^{+}\left(  \varepsilon\right)  \Bigr],\label{redA}
\end{eqnarray}
\begin{eqnarray}
&&\widetilde B_{\alpha m,\beta n}\left(  r,s_1,s_2\right)\notag\\
=&&\int\frac{d\varepsilon}{\pi}\operatorname{Re}  \frac{1}{(\varepsilon-s_{1}+i\eta)(\varepsilon-s_{2}-i\eta)}\Gamma_{\alpha m}\left(\varepsilon\right)  \Gamma_{\beta n}\left(  \varepsilon+r\right)\notag\\
&&\times \Bigl[  \Psi_{\beta n}^{+}\left(  \varepsilon+r\right)-\Psi_{\alpha m}^{+}\left(  \varepsilon\right)  \Bigr],\label{redB}
\end{eqnarray}
by the mathematical relations $f_{\alpha m}\left(\varepsilon_1\right)  \bar f_{\beta n}\left(  \varepsilon_2\right)=n_B(\varepsilon_1-\mu_{\alpha\beta}-\varepsilon_2+\mu_{\beta n})[f_{\beta n}\left(  \varepsilon_2\right)-f_{\alpha m}\left(  \varepsilon_1\right)]$, $f_{\alpha m}(\varepsilon)=\frac{1}{2}[1+\frac{i}{\pi}\Psi_{\alpha m}^+(\varepsilon)-\frac{i}{\pi}\Psi_{\alpha m}^-(\varepsilon)]$,
$\Psi_{\alpha m}^{\pm}\left(  \varepsilon\right)=\Psi\left(  \frac{1}{2}\pm i\frac{\varepsilon-\mu_{\alpha m}}{2\pi k_B T}\right)$, and $\left[\Psi_{\alpha m}^{+}\left(\varepsilon\right)\right]^\ast=\Psi_{\alpha m}^{-}\left(  \varepsilon^{\ast}\right)$. Here, $\Psi\left(z\right)$ is the digamma function possessing poles at $z=0,-1,-2,-3,...$; therefore, $\Psi_{\alpha m}^{+}\left(  \varepsilon\right)  $ and $\Psi_{\alpha m}^{-}\left(  \varepsilon\right)  $ have poles in the upper and lower complex half-planes, respectively. $\Psi\left(z\right)$ has the property $\Psi^{\ast}\left(z\right)=\Psi\left(z^{\ast}\right)$.

\begin{widetext}
We proceed to evaluate $\widetilde A_{\alpha m,\beta n}\left(  r,s\right)$ by contour integration. To avoid the poles of the $\Psi^+$ function appearing in Eq.~(\ref{redA}), we choose the closed semicircle with radius $R\rightarrow\infty$ in the lower complex half-plane as the contour, within which the functions $\Gamma_{\alpha m}(\varepsilon)=\frac{\Delta _{\alpha m}W_{\alpha m}^{2}}{\left( \varepsilon -\mu _{\alpha m}\right) ^{2}+W_{\alpha m}^{2}}$ and $\Gamma_{\beta n}(\varepsilon+r)=\frac{\Delta _{\beta n}W_{\beta n}^{2}}{\left( \varepsilon+r -\mu _{\beta n}\right) ^{2}+W_{\beta n}^{2}}$ each have a pole. The integral is given by the sum over the residues of the integrand at the three poles $s-i\eta$, $\mu_{\alpha m}-iW_{\alpha m}$, and $\mu_{\beta n}-r-iW_{\beta n}$,
\begin{eqnarray}
\widetilde{A}_{\alpha m,\beta n}\left(  r,s\right)&&=\frac{\Omega}{\eta}\frac{\Psi_{\beta n}^{+}\left(s-i\eta+r\right)  -\Psi_{\alpha m}^{+}\left(  s-i\eta\right)  }{\left[\left(  s-i\eta-\mu_{\alpha m}\right)  ^{2}+W_{\alpha m}^{2}\right]  \left[\left(  s-i\eta+r-\mu_{\beta n}\right)  ^{2}+W_{\beta n}^{2}\right]  }\notag\\
&&+\frac{\Omega}{W_{\alpha m}}\frac{\Psi_{\beta n}^{+}\left(\mu_{\alpha m}-iW_{\alpha m}+r\right)  -\Psi_{\alpha m}^{+}\left(  \mu_{\alpha m}-iW_{\alpha m}\right)  }{\left[  \left(  \mu_{\alpha m}-iW_{\alpha
m}-s\right)  ^{2}+\eta^{2}\right]  \left[  \left(  \mu_{\alpha m}-iW_{\alpha m}+r-\mu_{\beta n}\right)  ^{2}+W_{\beta n}^{2}\right]  }\notag\\
&&+\frac{\Omega}{W_{\beta n}}\frac{\Psi_{\beta n}^{+}\left(\mu_{\beta n}-iW_{\beta n}\right)  -\Psi_{\alpha m}^{+}\left(  \mu_{\beta n}-r-iW_{\beta n}\right)  }{\left[  \left(  \mu_{\beta n}-r-iW_{\beta
n}-s\right)  ^{2}+\eta^{2}\right]  \left[  \left(  \mu_{\beta n}-r-iW_{\beta n}-\mu_{\alpha m}\right)  ^{2}+W_{\alpha m}^{2}\right]  }, \label{contourintegral}
\end{eqnarray}
where $\Omega=\Delta_{\alpha m}W_{\alpha m}^{2}\Delta_{\beta n}W_{\beta n}^{2}$. As $\eta$ is a positive infinitesimal, the first term on the right-hand-side of Eq.~(\ref{contourintegral}) diverges. A standard regularization scheme \cite{PhysRevB.65.115332,PhysRevB.70.195107} curing this divergence is to expand the divergent term in powers of $\eta$ and then subtract the parts scaling as $1/\eta$ since they are indeed sequential tunneling processes and should be dropped to avoid double counting and, finally, take the limit $\eta\rightarrow 0^+$,
\begin{eqnarray}
\widetilde{A}_{\alpha m,\beta n}\left(  r,s\right)&&=\left[  \frac{2i\Omega\left(  s-\mu_{\alpha m}\right)  }{\left(  s-\mu_{\alpha m}\right)^{2}+W_{\alpha m}^{2}}+\frac{2i\Omega\left(  s+r-\mu_{\beta n}\right)  }{\left(s+r-\mu_{\beta n}\right)  ^{2}+W_{\beta n}^{2}}\right]  \frac{  \Psi_{\beta n}^{+}\left(  s+r\right)  -\Psi_{\alpha m}^{+}\left(  s\right)    }{\left[  \left(  s-\mu_{\alpha m}\right)^{2}+W_{\alpha m}^{2}\right]  \left[  \left(  s+r-\mu_{\beta n}\right)^{2}+W_{\beta n}^{2}\right]  }\notag\\
&&+\frac{\Omega}{2\pi k_{B}T}\frac{\Psi_{\beta n}^{\prime+}\left(  s+r\right)  -\Psi_{\alpha m}^{\prime+}\left(  s\right)  }{\left[\left(  s-\mu_{\alpha m}\right)  ^{2}+W_{\alpha m}^{2}\right]  \left[  \left(s+r-\mu_{\beta n}\right)  ^{2}+W_{\beta n}^{2}\right]  }\notag\\
&&+\frac{\Omega}{W_{\alpha m}}\frac{\Psi_{\beta n}^{+}\left(\mu_{\alpha m}-iW_{\alpha m}+r\right)  -\Psi_{\alpha m}^{+}\left(  \mu_{\alpha m}-iW_{\alpha m}\right)  }{\left(  \mu_{\alpha m}-iW_{\alpha m}-s\right)
^{2}\left[  \left(  \mu_{\alpha m}-iW_{\alpha m}+r-\mu_{\beta n}\right)^{2}+W_{\beta n}^{2}\right]  }\notag\\
&&+\frac{\Omega}{W_{\beta n}}\frac{\Psi_{\beta n}^{+}\left(\mu_{\beta n}-iW_{\beta n}\right)  -\Psi_{\alpha m}^{+}\left(  \mu_{\beta n}-r-iW_{\beta n}\right)  }{\left(  \mu_{\beta n}-r-iW_{\beta n}-s\right)
^{2}\left[  \left(  \mu_{\beta n}-r-iW_{\beta n}-\mu_{\alpha m}\right)^{2}+W_{\alpha m}^{2}\right]  },
\end{eqnarray}
where $\Psi^{'+}(z)$ denotes the first derivative of $\Psi^+(z)$. Along the same lines, one can obtain
\begin{eqnarray}
\widetilde{B}_{\alpha m,\beta n}\left(  r,s_{1},s_{2}\right)&&=\frac{i\Omega}{s_{2}-s_{1}}\frac{ \Psi_{\beta n}^{+}\left(  s_{1}+r\right)  -\Psi_{\alpha m}^{+}\left(  s_{1}\right)  }{\left[  \left(  s_{1}-\mu_{\alpha m}\right)  ^{2}+W_{\alpha m}^{2}\right]  \left[  \left(  s_{1}+r-\mu_{\beta n}\right)  ^{2}+W_{\beta n}^{2}\right]  }\notag\\
&&-\frac{i\Omega}{s_{2}-s_{1}}\frac{ \Psi_{\beta n}^{+}\left(  s_{2}+r\right)  -\Psi_{\alpha m}^{+}\left(  s_{2}\right)  }{\left[  \left(  s_{2}-\mu_{\alpha m}\right)  ^{2}+W_{\alpha m}
^{2}\right]  \left[  \left(  s_{2}+r-\mu_{\beta n}\right)  ^{2}+W_{\beta n}^{2}\right]  }\notag\\
&&+\frac{\Omega}{W_{\alpha m}}\frac{\Psi_{\beta n}^{+}\left(\mu_{\alpha m}-iW_{\alpha m}+r\right)  -\Psi_{\alpha m}^{+}\left(  \mu_{\alpha m}-iW_{\alpha m}\right)  }{\left[  \left(  \mu_{\alpha m}-iW_{\alpha m}
-s_{1}\right)  \left(  \mu_{\alpha m}-iW_{\alpha m}-s_{2}\right)  \right]\left[  \left(  \mu_{\alpha m}-iW_{\alpha m}+r-\mu_{\beta n}\right)^{2}+W_{\beta n}^{2}\right]  }\notag\\
&&+\frac{\Omega}{W_{\beta n}}\frac{\Psi_{\beta n}^{+}\left(\mu_{\beta n}-iW_{\beta n}\right)  -\Psi_{\alpha m}^{+}\left(  \mu_{\beta n}-r-iW_{\beta n}\right)  }{\left[  \left(  \mu_{\beta n}-r-iW_{\beta n}
-s_{1}\right)  \left(  \mu_{\beta n}-r-iW_{\beta n}-s_{2}\right)  \right]\left[  \left(  \mu_{\beta n}-r-iW_{\beta n}-\mu_{\alpha m}\right)^{2}+W_{\alpha m}^{2}\right]  }.\notag\\
\end{eqnarray}

\end{widetext}

%\bibliographystyle{apsrev4-1-etal-title}
%\bibliographystyle{apsrev4-1}
%\bibliography{ref-drag}


\end{document}